\documentclass{svjour3}                     
\smartqed  
\usepackage{graphicx}
\usepackage{bm}
\begin{document}

\title{Muon capture on light nuclei
\thanks{"Relativistic Description of Two- and Three-Body Systems in Nuclear
Physics", ECT*, October 19-13 2009}
}


\author{L.E.\ Marcucci    \and
        M. Piarulli
}

\institute{L.E.\ Marcucci \at
           Department of Physics, University of Pisa, and INFN-Pisa, \\
           56127 Pisa, Italy\\
           \email{laura.marcucci@edf.unipi.it}       
           \and
           M.\ Piarulli \at
           Department of Physics, Old Dominion University, \\
           Norfolk, VA 23529, USA
           \email{mpiar001@odu.edu}
}

\date{Received: date / Accepted: date}

\maketitle

\begin{abstract}

This work investigates 
the muon capture 
reactions $^2$H($\mu^-,\nu_\mu$)$nn$ and \\
$^3$He($\mu^-,\nu_\mu$)$^3$H and the contribution 
to their total capture rates arising from the axial two-body
currents obtained imposing the 
partially-conserved-axial-current (PCAC) hypothesis.
The initial and final $A=2$ and 3 
nuclear wave functions are obtained from the Argonne $v_{18}$ 
two-nucleon potential, 
in combination with the Urbana IX three-nucleon potential 
in the case of $A=3$. 
The weak current consists of
vector and axial components
derived in chiral effective field theory.  
The low-energy constant entering the
vector (axial) component is determined by
reproducting the isovector combination
of the trinucleon magnetic moment (Gamow-Teller
matrix element of tritium beta-decay).
The total capture rates are 393.1(8) s$^{-1}$ for $A=2$ and 1488(9) s$^{-1}$ 
for $A=3$, where the uncertainties arise from the adopted fitting procedure.

\end{abstract}

\section{Introduction}
\label{sec:intro}

There is a significant body of experimental and theoretical work
on muon captures in light nuclei, motivated by the fact that the
theoretical framework used to study these reactions is the same as
that used for weak capture reactions
of astrophysical interest, not accessible experimentally.
Muon captures, whose rates can be measured, can therefore provide 
a valuable test of this theoretical framework~\cite{SFII}.

Very recently~\cite{Mar10}, 
the muon capture reactions $^2$H($\mu^-,\nu_\mu$)$nn$ and 
$^3$He($\mu^-,\nu_\mu$)$^3$H have been studied
simultaneously in a consistent framework.
In particular, the initial and final $A=2$ and 3 
nuclear wave functions have been obtained from the 
Argonne $v_{18}$ (AV18)~\cite{Wir95} 
or the chiral N$^3$LO (N3LO)~\cite{Ent03} two-nucleon potential, 
in combination with, respectively, the Urbana IX (UIX)~\cite{Pud95} 
or chiral N$^2$LO (N2LO)~\cite{Nav07} 
three-nucleon potentials in the case of $A=3$. 
The weak current consists of polar- and axial-vector components. The
former are related to the isovector piece of the electromagnetic
current via the conserved-vector-current (CVC) hypothesis.
These and the
axial current have been derived within two
different frameworks, the standard nuclear
physics approach (SNPA), and chiral
effective field theory.
The first one goes beyond the impulse approximation, by including 
meson-exchange current contributions and terms arising
from the excitation of $\Delta$-isobar degrees of freedom.
The second approach includes two-body contributions
derived in 
heavy-baryon chiral perturbation theory 
within a systematic expansion, up to 
N$^3$LO~\cite{Par03,Son09}.
To be noticed that, since the transition operator matrix elements
are calculated using phenomenological wave functions, it should be
viewed as a hybrid chiral effective field theory approach (EFT*).
Both SNPA and EFT* frameworks have been used in studies
of weak $pp$ and $hep$ capture reactions in the energy regime relevant to
astrophysics~\cite{Par03,Sch98,Mar00}.
The only parameter in the SNPA 
nuclear weak current model is present in the axial current 
(the $N$-to-$\Delta$ axial coupling constant)
and is determined by fitting the experimental
Gamow-Teller matrix element in tritium $\beta$-decay (GT$^{\rm EXP}$).
The SNPA weak vector current, related
to the isovector electromagnetic current via CVC,  
reproduces the trinucleon magnetic moments to better than
1 \%~\cite{Mar10}. 
In the case of EFT*, three 
low-energy constants (LECs) appear: one in the axial-vector 
component, and two in the electromagnetic current. Of these, only 
one is relevant to the weak vector current, since the other 
appears in front of an isoscalar operator.
The corresponding coupling constants are parameters 
fixed to reproduce, respectively, 
GT$^{\rm EXP}$ and $A=3$ magnetic moments.
To be noticed that the EFT* currents are obtained performing the
Fourier transform from momentum- to coordinate-space with 
a Gaussian regulator characterized by a cutoff $\Lambda$, varied
between 500 and 800 MeV.
The total capture rates have been found to be 
392.0(2.3) s$^{-1}$ for $A=2$ and 
1484(13) s$^{-1}$  for $A=3$. 
The spread accounts for the model dependence, i.e., 
the dependence on the input Hamiltonian model, 
the model for the nuclear transition operator, and, in the EFT* 
calculation, the cutoff sensitivity.
This weak model dependence is a consequence of the procedure
adopted to constrain the weak current. These results
are in very good agreement with the experimental data, 
in particular with the very 
accurate measurement 
of Ref.~\cite{Ack98} for the total rate in muon capture on $^3$He.

The muon capture on deuteron has been studied 
also using the SNPA and the EFT* framework in Ref.~\cite{Ric10}. 
The SNPA retains two-body meson-exchange currents 
derived from the hard pion chiral Lagrangians of the 
$N\Delta\pi\rho\omega a_1$ system and are significantly
different from those of Ref.~\cite{Mar10}. On the other hand,
the EFT* currents are similar to those of Ref.~\cite{Mar10},
but two differences need to be remarked: (i) the LEC appearing in the
axial-vector component ($d_R$) is not fixed
to reproduce GT$^{\rm EXP}$, rather the doublet capture rate
calculated in SNPA; (ii) a term is added to the leading axial two-body
currents, in order to satisfy the partially-conserved-axial-current (PCAC)
hypothesis, as constructed in Ref.~\cite{Mos05} (called
there, and from now on, potential current). The calculated
SNPA values for the total capture rate are 
in the range of 416--430 s$^{-1}$,
depending on the potential model used, resulting in a 
model dependence much larger than in Ref.~\cite{Mar10}. 
It is also argued that ``omitting the potential current causes
an enhancement of the doublet transition rate $\Lambda_{1/2}$
by $\simeq$ 1\%''~\cite{Ric10}. 

In the present work we repeat the calculation of Ref.~\cite{Mar10},
in the EFT* approach, adding the potential currents as
in Ref.~\cite{Ric10}. We restrict our calculation to 
the AV18 and AV18/UIX potential models, and to 
a cutoff value of 600 MeV, 
since, as shown in Ref.~\cite{Mar10}, the dependence on these
inputs is less than 1 \%. We fit the $d_R$ coefficient to GT$^{\rm EXP}$, and 
consistently calculate the total rates
for muon capture on deuteron and $^3$He. The comparison with the results
of Ref.~\cite{Mar10} will give and indication of how significant are
the potential current contributions for these muon captures.

The paper is organized as follows: in Sec.~\ref{sec:thform} the theoretical
formalism used in the calculation is briefly reviewed. In Sec.~\ref{sec:wcur}
the EFT* model for the weak current is described, with the addition of
the potential currents. In Sec.~\ref{sec:res}, 
the results are presented and discussed, and some concluding remarks are
given.

\section{Theoretical formalism}
\label{sec:thform}
We briefly review the formalism used in the calculation for the muon
capture processes, discussed at length in Refs.~\cite{Mar10,Mar02}. 
The muon capture on deuteron
and $^3$He is induced by the weak interaction
Hamiltonian~\cite{Wal95}, 
$H_{W}={G_{V}\over{\sqrt{2}}} \int {\rm d}{\bf x}
\, l_{\sigma}({\bf x}) j^{\sigma}({\bf x})$, 
where $G_{V}$ is the Fermi coupling constant,
$G_{V}$=1.14939 $\times 10^{-5}$ GeV$^{-2}$~\cite{Har90},
and $l_\sigma$ and $j^\sigma$ are the leptonic and
hadronic current densities, respectively.  
The transition amplitude can be written as
\begin{eqnarray}
&&T_W (f,f_z;s_1,s_2,h_\nu) \equiv
\langle nn, s_1, s_2; \nu, h_\nu \,|\, H_W \,|\,
(\mu,d);f,f_z \rangle
\nonumber \\
&&\>\>\>\>\simeq {G_V \over \sqrt{2}} \psi_{1s}^{\rm av}
\sum_{s_\mu s_d}
\langle {1 \over 2}s_{\mu}, 1 s_d | f f_z \rangle\, 
l_\sigma(h_\nu,\,s_\mu)\, 
\langle \Psi_{{\bf p}, s_1 s_2}(nn) | j^{\sigma}({\bf q}) |
\Psi_d(s_d)\rangle \ , \label{eq:h2ffz}
\end{eqnarray}
for muon capture on deuteron, ${\bf p}$ being the $nn$ relative
momentum, and~\cite{Mar02}
\begin{eqnarray}
&&T_W (f,f_z;s^\prime_{3},h_\nu) \equiv
\langle ^3{\rm H}, s^\prime_{3}; \nu, h_\nu \,|\, H_W \,|\,
(\mu,^3\!{\rm He});f,f_z \rangle
\nonumber \\
&&\>\>\>\>\>\>\simeq {G_V \over \sqrt{2}} \psi_{1s}^{\rm av}
\sum_{s_\mu s_3}
\langle {1 \over 2}s_{\mu}, {1 \over 2} s_3 | f f_z \rangle\,
l_\sigma(h_\nu,\,s_\mu)\,
\langle \Psi_{^3{\rm H}}(s^\prime_{3}) | j^{\sigma}({\bf q}) |
\Psi_{^3{\rm He}} (s_3)\rangle \ , \label{eq:h3ffz}
\end{eqnarray}
for muon capture on $^3$He. In order to account for the hyperfine structure
in the initial system, the muon and deuteron or $^3$He
spins are coupled to states with total spin $f=1/2$ or 3/2 in the
deuteron case, and $f=0$ or 1 in the $^3$He case.
In Eqs.~(\ref{eq:h2ffz}) and~(\ref{eq:h3ffz}) we have defined 
with $s_\mu$ ($h_\nu$) the muon spin (muon neutrino helicity).
The Fourier transform of the nuclear weak current
has been introduced as
\begin{equation}
j^\sigma({\bf q})=\int {\rm d}{\bf x}\,
{\rm e}^{ {\rm i}{\bf q} \cdot {\bf x} }\,j^\sigma({\bf x})
\equiv (\rho({\bf q}),{\bf j}({\bf q}))
\label{eq:jvq} \>\>\>,
\end{equation}
with the leptonic momentum transfer ${\bf q}$ defined
as ${\bf q} = {\bf k}_\mu-{\bf k}_\nu \simeq -{\bf k}_\nu$,
${\bf k}_\mu$ and ${\bf k}_\nu$ being the muon and muon neutrino momenta.
The function $\psi_{1s}^{\rm av}$ has been introduced to take into account
the initial bound state of the muon in the atom and the charge
distribution of the nucleus. It is typically approximated
as~\cite{Wal95} $|\psi_{1s}^{\rm av}|^2 \,=\,
{(\alpha\, \mu_{\mu d})^3\over \pi}$ for muon capture on deuteron,
and~\cite{Mar02} $|\psi_{1s}^{\rm av}|^2 \,=\,
{\cal {R}}\,{(2\,\alpha\, \mu_{\mu ^3{\rm He}})^3\over \pi}$ 
for muon capture on $^3$He, where 
$\alpha$ is the fine structure constant ($\alpha=1/137$), 
$\mu_{\mu d}$ and $\mu_{\mu ^3{\rm He}}$
are the reduced masses of the $(\mu,d)$ and ($\mu,^3$He) systems, 
and the factor ${\cal {R}}$ approximately accounts
for the finite extent of the nuclear charge
distribution~\cite{Wal95} and is taken to be 0.98, as in Ref.~\cite{Mar10}.

In the case of muon capture on deuteron, the final state wave
function $\Psi_{{\bf p}, s_1 s_2}(nn)$ is expanded in partial waves, 
and the calculation is restricted
to total angular momentum $J\leq 2$ and orbital angular momentum 
$L\leq 3$, i.e., 
in a spectroscopic notation, to $^1S_0$, $^3P_0$, $^3P_1$, $^3P_2$--$^3F_2$ 
and $^1D_2$. 
Standard techniques~\cite{Mar00,Wal95} are
now used to carry out the multipole expansion
of the weak charge, $\rho({\bf q})$, and current,
${\bf j}({\bf q})$, operators. Details 
of the calculation can be found in Ref.~\cite{Mar10}.
Here we only note that all the contributing multipole operators 
selected by parity and angular momentum selection rules 
are included, as explained in Ref.~\cite{Mar10}.

The total capture rate for the two reactions under consideration
is then defined as
\begin{equation}
d\Gamma = 2\pi\delta(\Delta E) \overline{|T_W|^2} \times({\rm phase \, space})
\ ,
\label{eq:dgamma}
\end{equation}
where $\delta(\Delta E)$ is the energy-conserving $\delta$-function, 
and the phase space is 
$d{\bf p}\,d{\bf k}_\nu/(2\pi)^6$ 
for muon capture on deuteron and just
$d{\bf k}_\nu/(2\pi)^3$ for muon capture on $^3$He.
The following notation has been introduced: (i) for muon capture
on deuteron
\begin{equation}
\overline{|T_W|^2} = \frac{1}{2f+1}\sum_{s_1 s_2 h_\nu}\sum_{f_z}
|T_W(f,f_z;s_1,s_2, h_\nu)|^2 \ ,
\label{eq:hw2}
\end{equation}
and the initial hyperfine state has been fixed to be $f=1/2$; 
(ii) for muon capture on $^3$He
\begin{equation}
\overline{|T_W|^2} = \frac{1}{4}\, \sum_{s_3^\prime  h_\nu}\sum_{f  f_z}
|T_W(f,f_z;s^\prime_3, h_\nu)|^2 \ ,
\label{eq:hw3}
\end{equation}
and the factor 1/4 follows from assigning the same probability 
to all different hyperfine states. 

After carrying out the spin sums,
the differential rate for muon capture on deuteron 
($d\Gamma^D/dp$)
and the total rate for muon capture on $^3$He ($\Gamma_0$) are easily 
obtained, and their expressions can be found in Ref.~\cite{Mar10}.
In order to obtain the total rate $\Gamma^D$ for muon capture on deuteron,
$d\Gamma^D/dp$ is plotted versus $p$ and numerically integrated.

Bound and continuum wave functions for both two- and three-nucleon
systems entering in Eqs.~(\ref{eq:h2ffz}) and~(\ref{eq:h3ffz}) 
are obtained with the hyperspherical-harmonics (HH)
expansion method. This method, as implemented in the case of
$A=3$ systems, has been reviewed in considerable detail in a series
of recent publications~\cite{Kie08,Viv06,Mar09}. We have used the 
same method 
in the context of $A=2$ systems, for which of course
wave functions could have been obtained by direct solution
of the Schr\"odinger equation. A detailed discussion for the
$A=2$ wave functions is given in Ref.~\cite{Mar10}.

\section{The nuclear weak current operator}
\label{sec:wcur}

The chiral effective field theory weak current transition operator 
is taken from Refs.~\cite{Par03} and~\cite{Son09}, as reviewed
in Ref.~\cite{Mar10}. It is 
derived in covariant perturbation theory based on the heavy-baryon
formulation of chiral Lagrangians by retaining corrections 
up to N$^3$LO. 
The one-body operators are those listed in 
Eqs.~(17) of Ref.~\cite{Par03} and~(4.13)--(4.14) of Ref.~\cite{Mar10}.
The vector charge and axial current operators retain terms up to $1/m^2$,
while the axial charge and vector current operators retain terms
up to $1/m^3$, $m$ being the nucleon mass. Both $1/m^2$ and $1/m^3$ 
contributions arise when the non-relativistic reduction of the 
single-nucleon covariant current is pushed to next-to-leading order.
The two-body vector currents are obtained from the two-body
electromagnetic currents via CVC. These are
decomposed into four terms~\cite{Son09}: 
the soft one-pion-exchange ($1\pi$) term, vertex corrections to the one-pion
exchange ($1\pi C$), the two-pion exchange ($2\pi$), and a contact-term
contribution. Their explicit expressions can be found in 
Ref.~\cite{Son09}. All the
$1\pi$, $1\pi C$ and $2\pi$ contributions contain low-energy constants
estimated using resonance saturation arguments, and
Yukawa functions obtained by performing the Fourier transform from 
momentum- to coordinate-space with a Gaussian regulator characterized
by a cutoff $\Lambda$. Here, as discussed above,
we fixed the value of $\Lambda=600$ MeV.
The contact-term electromagnetic contribution
is given as sum of two terms, isoscalar and isovector, each one 
with a coefficient in front ($g_{4S}$ and $g_{4V}$) fixed to
reproduce the experimental values of triton and $^3$He magnetic 
moments. For the AV18/UIX Hamiltonian model, with $\Lambda=600$ MeV, 
$g_{4S}=0.55(1)$ and $g_{4V}=0.793(6)$, the error being due to 
numerics~\cite{Mar10}. Note that only the isovector contribution
is of interest here, but anyway it turns out to be negligible.

The two-body axial current operator consists of two contributions:
a one-pion exchange term and a (non-derivative) two-nucleon 
contact-term. The explicit expression for the contact term
can be found in Ref.~\cite{Par03}. Here we review the one-pion
exchange term, since we add, in accordance with Ref.~\cite{Mos05},
the potential current contributions. Therefore, in momentum-space,
the one-pion exchange term reads:
\begin{eqnarray}
{\bf j}_{ij}^{\pi}({\bf q}; A)&=&
\frac{g_A}{2 m f_\pi^2}\Bigl[({\bm \tau}_i\times{\bm \tau}_j)^a
\Bigl[\frac{\rm i}{2}\,(1-g_A^2)\,\frac{{\bf p}_i+{\bf p}_i^\prime}{2}
+\big(\frac{1}{4}+{\hat c}_4\big)\,{\bm\sigma}_i\times{\bf k}_j
\nonumber \\
&&\>\>\>\>\>\>\>\>\>\>\>\>\>\>\>\>\>\>\>\>\>\>\>\>\>\>\>\>\>\>\>\>
+(\frac{1+c_6+g_A^2}{4})\,{\bm\sigma}_i\times{\bf q}\Bigr]
+2{\hat c}_3{\bm \tau}_j^a{\bf k}_j
\nonumber \\
&&\>\>\>\>\>\>\>\>\>\>\>
-\frac{g_A^2}{4}{\bm\tau}_j^a\,\Bigl({\bf q}+{\rm i}{\bm \sigma}_i\times
({\bf p}_i+{\bf p}_i^\prime)\Bigr)\Bigr]
\frac{{\bm\sigma}_j\cdot{\bf k}_j}{m_\pi^2+{\bf k}_j^2} + i\leftrightarrow j
\ .
\label{eq:jwa}
\end{eqnarray}
where ${\bf k}_{i,j}={\bf p}_{i,j}^\prime-{\bf p}_{i,j}$, 
with ${\bf p}_{i,j}$ and ${\bf p}_{i,j}^\prime$ being the initial
and final single nucleon momenta, ${\bf q}={\bf k}_i+{\bf k}_j$, 
$g_A=1.2654$ is the axial-vector coupling constant, and $f_\pi=93$ MeV
is the pion decay constant. 
The values used for the coupling constants ${\hat c}_3$, ${\hat c}_4$, 
and $c_6$, as obtained from $\pi N$ data, 
are ${\hat c}_3=-3.66$, ${\hat c}_4=2.11$ and $c_6=5.83$~\cite{Par03}.
The terms proportional to $g_A^2$ in Eq.~(\ref{eq:jwa})
are the potential currents. They
are the same as in Eq.~(21) of Ref.~\cite{Mos05} or
Eq.~(A.17) of Ref.~\cite{Ric10}.
The low-energy constant $d_R$, 
determining the strength of the contact-term two-body axial
contribution, has been fixed 
by reproducing GT$^{\rm EXP}$, finding $d_R=1.54(8)$. This value should
be compared with the corresponding one given in Ref.~\cite{Mar10}
(see Table V), $d_R=1.75(8)$. The difference between these two values
of $\simeq$ 13 \% is due to the presence of the potential currents
and is comparable with that of Ref.~\cite{Ric10}.

\section{Results}
\label{sec:res}

We present in Table~\ref{tab:res} the results for the total rates
of muon capture on deuteron, in the doublet hyperfine state ($\Gamma^D$), 
and on $^3$He ($\Gamma_0$). 
The deuteron, $nn$, $^3$He and $^3$H wave functions have been
calculated with the  
AV18~\cite{Wir95} two- and, when necessary, UIX~\cite{Pud95} 
three-nucleon interactions. 
The model for the  nuclear weak transition operator has been 
presented in Sec.~\ref{sec:wcur}. 
We compare our results with those of Ref.~\cite{Mar10}, 
obtained with the same Hamiltonian model and cutoff $\Lambda$, 
but without the two-body potential currents elaborated
in Ref.~\cite{Mos05} and discussed in Sec.~\ref{sec:wcur}. 
Note that here $d_R=1.54(8)$, while in
Ref.~\cite{Mar10} $d_R=1.75(8)$.
From inspection of the table, we conclude that the two calculations
are in remarkable agreement with each other: the differences in $\Gamma_0$ 
and $\Gamma^D$ are $\leq 0.1$ \%, well below the theoretical
uncertainties. The largest difference, of the
order of 2 \%, is in the $^3P_0$ partial wave contribution
to $\Gamma^D$. However, when all the partial
wave contributions are summed up, the difference in $\Gamma^D$ 
returns well below the 1 \% level.

In conclusion, we have studied the potential currents 
dictated by PCAC, as elaborated in Refs.~\cite{Ric10,Mos05}, 
and we have found that their contributions to the total rates
of muon capture on deuteron and $^3$He are tiny. 
This result is a consequence of the procedure 
adopted to constrain the weak current. 
Finally, we expect that the potential currents 
will give tiny contributions also in 
weak capture reactions of astrophysical interest and in those
processes whose momentum transfer is small.

\begin{table}
\caption{Total rate for muon capture on deuteron and $^3$He, in  
s$^{-1}$. In the $A=2$ case, 
the different partial wave contributions are indicated. The
numbers among parentheses indicate the theoretical uncertainties
arising from the adopted fitting procedures.
Such uncertainty is not indicated when 
less than 0.1 s$^{-1}$. 
The AV18 and AV18/UIX interactions have been used to calculate the 
$A=2$ and $A=3$ wave functions.
The corresponding results of Ref.~\protect\cite{Mar10} are
also listed.}
\label{tab:res}       
\begin{tabular}{llllllll|l}
\hline\noalign{\smallskip}
& $^1S_0$ & $^3P_0$ & $^3P_1$ 
& $^3P_2$ &  $^1D_2$ & $^3F_2$ 
& $\Gamma^D$ & $\Gamma_0$ \\ 
\noalign{\smallskip}\hline\noalign{\smallskip}
Present work & 250.1(8) & 20.2 & 46.1 & 71.3 & 4.5 & 0.9 & 393.1(8) & 1488(9) \\
Ref.~\protect\cite{Mar10} &
   250.0(8) & 19.8 & 46.3 & 71.1 & 4.5 & 0.9 & 392.6(8) & 1488(9) \\
\noalign{\smallskip}\hline
\end{tabular}
\end{table}


\begin{thebibliography}{}
%
%
\bibitem{SFII}
E.G.\ Adelberger {\it et al.},
arXiv:1004.2318
%
\bibitem{Mar10}
L.E.\ Marcucci {\it et al.},
arXiv:1008.1172
%
\bibitem{Wir95} R.B.\ Wiringa, V.G.J.\ Stoks, and R.\ Schiavilla,
                Phys.\ Rev.\ C {\bf 51}, 38 (1995)
%
\bibitem{Ent03} D.R.\ Entem and R.\ Machleidt, 
                Phys.\ Rev.\ C {\bf 68}, 041001 (2003)
%
\bibitem{Pud95} B.S.\ Pudliner {\it et al.},
                Phys.\ Rev.\ Lett.\ {\bf 74}, 4396 (1995)
%
\bibitem{Nav07} P.\ Navr\'atil,
                Few-Body Syst.\ {\bf 41}, 117 (2007)
%
\bibitem{Par03}
T.-S.\ Park {\it et al.}, Phys.\ Rev.\ C {\bf 67}, 055206 (2003)
%
\bibitem{Son09}Y.-H.\ Song, R.\ Lazauskas, and T.-S.\ Park,
Phys.\ Rev.\ C {\bf 79}, 064002 (2009)
%
\bibitem{Sch98}
R.\ Schiavilla {\it et al.}, Phys.\ Rev.\ C {\bf 58}, 1263 (1998)
%
\bibitem{Mar00}
L.E.\ Marcucci {\it et al.}, Phys.\ Rev.\ Lett.\ {\bf 84}, 5959 (2000);
Phys.\ Rev.\ C {\bf 63}, 015801 (2000)
%
\bibitem{Ack98}
P.\ Ackerbauer {\it et al.}, Phys.\ Lett.\ B {\bf 417}, 224 (1998)
%
\bibitem{Ric10}
P.\ Ricci, E.\ Truhlik, B.\ Mosconi, and J.\ Smejkal,
Nucl.\ Phys.\ A {\bf 837}, 110 (2010)
%
\bibitem{Mos05}
B.\ Mosconi, P.\ Ricci, and E.\ Truhl\'ik,
Eur.\ Phys.\ J.\ A {\bf 25}, 283 (2005)
%
\bibitem{Mar02}
L.E.\ Marcucci, R.\ Schiavilla, A.\ Kievsky, and M.\ Viviani,
Phys.\ Rev.\ C {\bf 66}, 054003 (2002)
%
\bibitem{Wal95} 
J.D.\ Walecka, {\it Theoretical Nuclear and Subnuclear Physics},
Oxford University Press, New York (1995)
%
\bibitem{Har90} 
J.C.\ Hardy {\it et al.},
Nucl.\ Phys.\ A {\bf 509}, 429 (1990)
%
\bibitem{Kie08} 
A.\ Kievsky {\it et al.}, 
J.\ Phys.\ G: Nucl.\ Part.\ Phys.\ {\bf 35}, 063101 (2008)
%
\bibitem{Viv06} 
M.\ Viviani {\it et al.},
Few-Body Syst.\ {\bf 39}, 159 (2006)
%
\bibitem{Mar09} 
L.E\ Marcucci {\it et al.},
Phys.\ Rev.\ C {\bf 80}, 034003 (2009)
%
\end{thebibliography}


\end{document}